\documentclass[12pt]{article}
\usepackage{amsmath,amssymb}

\usepackage{graphicx}
\usepackage{wrapfig}

\usepackage{hyperref}
\usepackage{cite}

\def\beq{\begin{eqnarray}}
\def\eeq{\end{eqnarray}}

\def\la{\langle }
\def\ra{\rangle }

\newcommand{\Tr}{\,\mathrm{Tr}\,}            

\newcommand{\be}{\begin{equation}}
\newcommand{\ee}{\end{equation}}
\newcommand{\bea}{\begin{eqnarray}}
\newcommand{\eea}{\end{eqnarray}}
\newcommand{\bg}{\begin{gather}}

\newcommand{\bseq}{\begin{subequations}}
\newcommand{\eseq}{\end{subequations}}

\renewcommand{\tanh}{\mathop{\rm th}\nolimits}

\renewcommand{\ln}{\mathop{\rm ln}\nolimits}

\def\tr{\hbox{Tr}}

\def\be{\begin{eqnarray}}
\def\ee{\end{eqnarray}}
\def\lb{\label}

\newcommand{\nn}{\nonumber}
\newcommand{\p}{\partial}
\begin{document}

\title{\textbf{Holographic calculation of boundary terms \\
in conformal anomaly }}

\vspace{2cm}
\author{ \textbf{  Amin Faraji Astaneh$^{1}$ and  Sergey N. Solodukhin$^2$ }} 
\date{}
\maketitle
\begin{center}
\hspace{-0mm}
  \emph{$^1$ School of Particles and Accelerators,\\ Institute for Research in Fundamental Sciences (IPM),}\\
\emph{ P.O. Box 19395-5531, Tehran, Iran}
 \end{center}
\begin{center}
  \hspace{-0mm}
  \emph{ $^2$ Laboratoire de Math\'ematiques et Physique Th\'eorique  CNRS-UMR
7350, }\\
  \emph{F\'ed\'eration Denis Poisson, Universit\'e Fran\c cois-Rabelais Tours,  }\\
  \emph{Parc de Grandmont, 37200 Tours, France}
\end{center}



\begin{abstract}
\noindent { In the presence of boundaries the integrated conformal anomaly is modified by the boundary terms
so that the anomaly is non-vanishing in any (even or odd) dimension. The boundary terms are due to extrinsic
curvature whose exact structure in $d=3$ and $d=4$ has recently been identified. In this note we present a holographic
calculation of those terms in two different prescriptions for  the holographic description of the boundary CFT.
We stress the role of supersymmetry when discussing the holographic description of ${\cal N}=4$ SYM on a $4$-manifold with boundaries.  
}
\end{abstract}

\vskip 2 cm
\noindent
\rule{7.7 cm}{.5 pt}\\
\noindent 
\noindent
\noindent ~~~ {\footnotesize e-mails:\ faraji@ipm.ir ,  Sergey.Solodukhin@lmpt.univ-tours.fr}


\newpage

\section{ Introduction}
The quantum effective action,  especially its UV divergent part, on manifolds with boundaries has been studied for many years \cite{Dowker:1989ue}.
The heat kernel technique,  relevant to this study for free fields of various spin, on manifolds with boundaries has been reviewed 
in \cite{Vassilevich:2003xt}.  These studies have been  recently revisited  that has resulted in an important observation that
the integrated conformal anomaly should be modified by the boundary terms if the manifold in question has boundaries 
\cite{Herzog:2015ioa},  \cite{Fursaev:2015wpa},   \cite{Solodukhin:2015eca}. Remarkably, the boundary anomaly is present for any (even and odd) dimension $d$.
This is drastically different from the local form of the anomaly which is present only in even dimensions \cite{Capper:1974ic}.
The boundary terms represent certain conformal invariants  constructed from the bulk Riemann curvature, intrinsic curvature on the boundary
and the extrinsic curvature.  The number of such invariants  rapidly grows with dimension $d$ and as for now there is no a complete classification of these 
boundary invariants. Formulating such a classification, by analogy with the one given in   \cite{Deser:1993yx} for the bulk conformal invariants, is an interesting 
open problem. For some recent progress in this direction see \cite{Glaros:2015pfa}.

Holography plays an important role in the study of strongly coupled conformal field theories.  It provides a purely geometric way of computing the 
important characteristics of the conformal theories. Some characteristics, such as the conformal anomalies, are protected by the non-renormalization theorems
and can be alternatively computed in the limit when the interaction is switched off.  The holographic calculation of the local conformal anomaly
in various dimensions was done in \cite{Henningson:1998gx}. In $d=4$ case the conformal field theory in question is ${\cal N}=4$ superconformal 
Yang-Mills theory. The holographic study thus provides us with certain important information, otherwise unavailable,  on this strongly coupled theory.

According to the AdS/CFT correspondence, the conformal field theory defined on manifold ${\cal M}_d$ which has a holographic dual is equivalent to a supergravity theory on
$(d+1)$-dimensional asymptotically Anti-de Sitter spacetime, $AdS_{d+1}$ whose asymptotic boundary is ${\cal M}_d$.
If ${\cal M}_d$ itself has a boundary $\partial {\cal M}_d$ this correspondence should be reformulated. Part of this reformulation is a prescription how $\partial{\cal M}_d$
is extended into the bulk of the Anti-de Sitter spacetime. This extension, we shall call it hypersurface ${\cal S}$ is such that its boundary is $\partial{\cal S}=\partial {\cal M}_d$.
On the other hand ${\cal S}$ is yet another component, additional to ${\cal M}_d$ of the boundary of the Anti-de Sitter spacetime $AdS_{d+1}$\footnote{We notice, however, the important difference between ${\cal M}_d$ and ${\cal S}$: ${\cal M}_d$ is conformal boundary at infinity that can be reached by a 
massive particle in infinite time while surface ${\cal S}$ can be reached in finite time from anywhere in the bulk. We thank K. Skenderis and M. Taylor for this remark.}.
We review the prescriptions available in the literature of how to define the hypersurface ${\cal S}$ and point out the certain subtleties.
We then suggest our own prescription: define ${\cal S}$ as a minimal surface. In all these prescriptions the $(d+1)$-dimensional gravitational action 
is modified by adding   certain boundary terms. In the prescription of Takayanagi et al. it is the Gibbons-Hawking term on $\cal S$.
In the prescription that we propose one has to add the volume of the surface $\cal S$. We then calculate holographically the boundary conformal anomaly 
in both prescriptions using same method as in \cite{Henningson:1998gx} by singling out the logarithmic term in the bulk gravitational action.
 For the minimal surface prescription 
 we find that if $d$ is odd then the logarithmic term originates from the volume of ${\cal S}$ while if $d$ is even it comes entirely from the bulk integral.
 The latter is due to  the fact that  the bulk integration in certain direction terminates at  surface ${\cal S}$  and thus the integral is affected by the shape
 of $\cal S$ which in turn  contains information on the co-dimension two boundary ${\cal M}_d$ and its extrinsic curvature $k$.
 This observation goes in parallel with the fact that for even $d$ the boundary anomaly contains odd powers of extrinsic curvature $k$ while 
 for odd $d$ the powers of $k$ in the anomaly are even. Clearly, the volume of a minimal hypersurface $\cal S$ can not contain\footnote{We thank R. Myers for pointing this out to us.}
  information on the direction of the normal vector to ${\cal M}_d$. On the other hand,
the bulk integration automatically picks the  outward normal to $\cal S$ and respectively to $\partial{\cal M}_d$. In the case when the boundary ${\cal S}$ is non-minimal
both the bulk integral and the area of the boundary produce some logarithmic terms.
 
 In the present note, for simplicity, we concentrate on $d=3$ and $d=4$ cases in which the boundary terms in anomaly are simple.
 Also, to make the computations simple we consider first the case when  manifold  ${\cal M}_d$ is flat so that  it is sufficient to only keep track  of the extrinsic curvature terms in the anomaly. The case of curved metric on ${\cal M}_4$ is fully treated  in section 4.5.
Throughout the paper extrinsic curvature $k$ is defined with respect to the outward normal vector.

\section{Boundary terms in conformal anomaly in $d = 3$ and $d = 4$ dimensions}

\subsection{Conformal invariant boundary conditions.}
Formulating the conformal field theory on a manifold with boundaries we should be sure that the boundary conditions to be imposed on the fields
do not break the conformal invariance. For  a field of spin $s$ it can be a combination of the Dirichlet boundary condition and the Neumann, or more generally Robin,
boundary condition.

For a conformal scalar field in $d$ dimensions there are two boundary conditions which are conformally invariant,
\be
&&{\rm Dirichlet\ \ b.\ c.}: \  \  \  \  \phi |_{\partial{\cal M}_d}=0\, , \nonumber \\
&&{\rm Robin \  \  b.\ c.}: \ \ \  \  \  \  \ (\partial_N+\frac{(d-2)}{2(d-1)}k)\phi |_{\partial{\cal M}_d}=0\, ,
\lb{-2}
\ee
where $k$ is the trace of extrinsic curvature of ${\partial{\cal M}_d}$. If the boundary is minimal then $k=0$ and the Robin boundary condition becomes 
the Neumann one.

For a massless Dirac fermion in dimension $d=4$ the conformal boundary condition is a mixed one: on impose Dirichlet boundary condition on a 
half of component of the spinor $\psi$ and the Robin type boundary condition on the other half,
\be
\Pi_- \psi |_{\partial{\cal M}}=0\, ,  \  \  \  (\nabla_N+K/2)\Pi_+\psi |_{\partial{\cal M}}=0\, ,
\lb{-1}
\ee
where $\Pi_\pm=\frac{1}{2} (1 \pm +i\gamma_*N^\mu \gamma_\mu)$, $N^\mu$ is normal vector and $\gamma_*$  is a chirality gamma matrix.

For a gauge field $A_\mu$ there are two boundary conditions which are manifestly gauge and conformal invariant,
\be
&&{\rm absolute\ \ b.\ c.}:    \  \  \  N^\mu F_{\mu\nu}=0\, , \nonumber \\
&&{\rm relative\  \  b.\ c.}: \  \  \    N^\mu F^{*}_{\mu\nu}=0\, ,
\lb{0}
\ee
where $F_{\mu\nu}=\partial_\mu A_\nu-\partial_\nu A_\mu$  is the strength of gauge field and $ F^*_{\mu\nu}$ is its Hodge dual.
In the Lorentz gauge each of these conditions reduces to a combination of the Dirichlet and Robin boundary conditions, see  \cite{Vassilevich:2003xt}.

\subsection{Conformal anomaly in dimension  $d=3$}
 In this case there are no bulk terms in the anomaly. The whole contribution comes from the boundary.   There are two possible  boundary terms which are conformally invariant: the Euler number of the boundary and the trace of  square of the trace-free extrinsic curvature, $\hat{k}_{ij}=k_{\ij}-\frac{1}{2}\gamma_{ij}k$, here we use the projection on the boundary so that indexes $i,j$ are along the two-dimensional surface.
Thus in this case the possible form of the anomaly is  \cite{Solodukhin:2015eca},  \cite{Fursaev:2016inw}
\be
\int_{{\cal M}_3}\la T\ra =\frac{c_1}{96}\chi[\partial{\cal M}_3]+\frac{c_2}{256\pi}\int_{\partial{\cal M}_3}\Tr\hat{k}^2\, ,
\lb{1}
\ee
where $\chi[\partial{\cal M}_3]=\frac{1}{4\pi}\int_{\partial{\cal M}_3} \hat{R}$ is the Euler number of the boundary, $\hat{R}$ is the  intrinsic scalar curvature of the boundary metric.
For a conformal scalar we find $c_1=-1$ and $c_2=1$ for the Dirichlet boundary condition and $c_1=1$ and $c_2=1$ for the conformal Robin condition, $(\partial_n+k/4)\phi|_{\partial{\cal M}_3}=0$. 

If manifold ${\cal M}_3$ is flat then the intrinsic curvature is related to the extrinsic curvature due to the Gauss-Codazzi relations
\be
\hat{R}=k^2-\tr k^2\, ,  \  \  k=\tr k\, .
\lb{2}
\ee
Since $\tr\hat{k}^2=\tr k^2-\frac{1}{2} k$ the anomaly (\ref{1}) can be expressed in terms of the extrinsic curvature only
\be
\int_{{\cal M}_3}\la T\ra=\frac{1}{256\pi}\int_{\partial{\cal M}_3} \left((c_2-\frac{2}{3}c_1)\tr k^2+(\frac{2}{3}c_1+\frac{c_2}{2})k^2\right)\, .
\lb{3}
 \ee
\bigskip
If $c_2=\frac{2}{3}c_1$ then $\tr k^2$ drops out in (\ref{3}). As we will see this is exactly what happens in the holographic calculation.

\subsection{Conformal anomaly in $d=4$}
The integrated conformal anomaly in four dimensions takes the following form,
\be
\int_{{\cal M}_4}\la T\ra=-\frac{a}{180}\chi[{\cal M}_4]+\frac{1}{1920\pi^2}\left(\int_{{\cal M}_4} b\Tr W^2-8b_1\int_{\partial{\cal M}_4}W^{\mu\nu\alpha\beta}n_\mu n_\beta \hat{k}_{\nu\alpha}\right)+\frac{c}{280\pi^2}\int_{\partial{\cal M}_4} \Tr\hat{k}^3\, ,
\lb{4}
\ee
where  $\hat{k}_{ij}=k_{ij}-\frac{1}{3}\gamma_{ij}k$ is traceless part of extrinsic curvature and the topological Euler number
\be
&&\chi[{{\cal M}_4}]=\frac{1}{32\pi^2}\int_{{\cal M}_4} (R_{\alpha\beta\mu\nu}R^{\alpha\beta\mu\nu}-4R_{\mu\nu}R^{\mu\nu}+R^2)\nonumber \\
&&-\frac{1}{4\pi^2}\int_{\partial{\cal M}_4}(-k^{\mu\nu}R_{n\mu n\nu}+k^{\mu\nu}R_{\mu\nu}+kR_{nn}-
\frac{1}{2}kR -\frac{1}{3}k^3+k\Tr k^2-\frac{2}{3}\Tr k^3)\, , 
\lb{5}
\ee
and $\Tr W^2=R_{\alpha\beta\mu\nu}R^{\alpha\beta\mu\nu}-2R_{\mu\nu}R^{\mu\nu}+\frac{1}{3}R^2$ is the square of the Weyl tensor.
In the normalization  used in eq.(\ref{3}) a scalar field has $a=b=1$. Notice the appearance of a boundary conformal charge $b_1$.
The direct calculation for free fields of spin $s=0,1/2 , 1$ shows that $b_1=b$. An argument why it should be so, based on variational principle applied to the integrated anomaly,
is given in \cite{Solodukhin:2015eca}.

If ${\cal M}_4$ is flat then the bulk terms in (\ref{2})-(\ref{3}) disappear and there remain only boundary terms expressed in terms of the extrinsic curvature,
\be
\int_{{\cal M}_4}\la T\ra=\frac{1}{\pi^2}\int_{\partial{\cal M}_4}\left(\frac{a}{720}(-\frac{1}{3}k^3+k\tr k^3-\frac{2}{3}\tr k^3)+\frac{c}{280}(\tr k^3-k\tr k^2+\frac{2}{9}k^3)\right)\, ,
\lb{6}
\ee
where we used that
\be
\tr\hat{k}^3=\tr k^3-k\tr k^2 +\frac{2}{9}k^3\, .
\lb{7}
\ee
Fursaev has computed the values of the boundary charges $b_1$ and $c$ for free fields \cite{Fursaev:2015wpa}.
They are listed below together with values of conformal charge $a$ for fields of different spin:
\be
&&{\rm real \ \ scalar:} \  \  \  \  \  \  \  \  a=1\,  ,  \  \  b_1=b=1\, , \   \ c=1 \ \  \  \ {\rm (Dirichlet \, b.\, c.)}\, , \\
&&{\rm real \ \ scalar:}\  \  \  \  \  \  \  \  a=1\,  , \  \  b_1=b=1 \, , \ \ c=\frac{7}{9} \ \  \  \ {\rm (Robin \, b.\, c.)}\, , \nonumber \\
&&{\rm Dirac \ \ fermion:}\ \  \  a=11\, , \  \  b_1=b=6\, , \ \  c=5, \ \  \ \ {\rm (mixed \, b.\, c.)}\, , \nonumber \\
&&{\rm gauge \ \ boson:} \ \ \  \  \    a=62\, , \  \   b_1=b=12\, ,  \  \ c=8\ \  \  \ {\rm (absolute\ \ or \ \ relative  \, b.\, c.)}\, .\nonumber
 \lb{8}
 \ee
We see that only $c$ charge  due to scalars appears to be sensitive to the  boundary conditions.
Using these values we can now compute the anomaly for a multiplet consisting $n_s^D$ scalars satisfying the Dirichlet boundary condition,
$n_s^R$ scalars satisfying the conformal Robin boundary conditions, $n_f$ massless Dirac fermions and $n_v$ gauge bosons.

\subsection{Conformal anomaly in $d=4$:  ${\cal N}=4$  $SU(N)$ super Yang-Mills multiplet}
In $d=4$, the conformal field theory which is holographically dual to the supergravity on $AdS_5$ is  ${\cal N}=4$  $SU(N)$ superconformal Yang-Mills theory.
The corresponding free field multiplet consists of $n_s=6$ scalars, $n_f=2$ Dirac fermions and $n_v=1$ gauge bosons, each field in the adjoint representation of $SU(N)$.
As we see from (\ref{8}) only scalars are sensitive to the choice of boundary conditions. The total number $n_s^D+n_s^R=6$ is fixed. Introducing 
$\Delta n=n^D_s-n^R_s$ we find,
\be
a=90(N^2-1)\, , \ \ \ b=b_1=30(N^2-1)\, , \ \ \ c=(\frac{70}{3}+\frac{1}{2}\Delta n)(N^2-1)\, .
\lb{9}
\ee
and hence the integral anomaly  is (we focus only on the boundary terms)
\be
\int_{{\cal M}_4}\la T\ra_{SYM}=\frac{(N^2-1)}{24\pi^2}\int_{\partial{\cal M}_4}[\frac{3}{2}(k^{\mu\nu}+k n^\mu n^\nu-\frac{2}{3}kg^{\mu\nu})R_{\mu\nu}
+(k\tr k^2-\frac{5}{9}k^3)+\frac{3\Delta n}{70}\tr \hat{k}^3]\, .
\lb{10}
\ee
It is well known that the Riemann tensor does not appear in the local conformal anomaly in ${\cal N}=4$ superconformal gauge theory so that
the anomaly vanishes in Ricci flat spacetime.  
We notice that  as well the Riemann tensor cancels in the boundary term (\ref{10}). So that the boundary term in the anomaly in Ricci flat spacetime is the same as in
Minkowski spacetime, provided the boundary is characterized by same extrinsic curvature. This property of the anomaly is not sensitive  to the choice of the conformal invariant boundary conditions imposed on the fields.
Additionally, we see that the last term in (\ref{10}), which is sensitive to the choice of the boundary conditions,   disappears if $n_s^D=n^R_s=3$ so that the term $\tr k^3$ drops out from the anomaly. In Ricci flat spacetime we have then
\be
\int_{{\cal M}_4}\la T\ra_{SYM}=\frac{(N^2-1)}{24\pi^2}\int_{\partial{\cal M}_4}(k\tr k^2-\frac{5}{9}k^3)\, .
\lb{11}
\ee
In this case one imposes the Dirichlet boundary condition on a half of scalars and the Robin boundary condition (which is a modification of the Neumann condition)
on the other half of scalars.  The presence of boundaries breaks the Lorentz symmetry and respectively the supersymmetry.  Some part of supersymmetry however can be
preserved if the boundary conditions are chosen appropriately. As was shown in \cite{Gaiotto:2008sa}, the condition that the preserved supersymmetry
in ${\cal N}=4 $ superconformal theory  is maximal  is precisely the conditions that $n_s^D=n_s^R$. 
In this case the boundaries preserve $1/2$ of supersymmetry. There is less supersymmetry if  $n_s^D\neq n_s^R$. 

The  charges in the conformal anomaly of a superconformal gauge theory are believed to be protected due to the non-renormalization theorems, as those proven in 
\cite{Petkou:1999fv},  so that they
are the same  for free field multiplet  and in the strong coupling regime accessible  holographically for $N\gg 1$.   It is an interesting question  whether these theorems can be extended to include the boundary charges
$b_1$ and $c$.  Validity of these theorems in the presence of boundaries  has not been analyzed to the best of our knowledge.

\section{Holographic prescriptions for BCFT}

\subsection{Takayanagi's prescription}
 The existing proposal for the holographic description of a boundary CFT is due to Takayanagi \cite{Takayanagi:2011zk}.
His prescription consists in adding a Gibbons-Hawking term as well as a boundary cosmological constant $T$ on the boundary $\cal S$ to the $(d+1)$-dimensional
bulk gravitational action,
\be
W^{T}_{gr}=-\frac{1}{16\pi G}\int_{{AdS}_{d+1}}(R-2\Lambda)-\frac{1}{8\pi G}[\int_{{\cal M}_d}K+\int_{{\cal S}_d}(K+T)]\, ,
\lb{12}
\ee
where $K$ is extrinsic curvature of co-dimension one boundary, either ${\cal M}_d$ or ${\cal S}_d$.
Variation with respect to boundary metric $\gamma_{ij}$ will give us the following equation
\be
K_{ij}-\gamma_{ij}K-T \gamma_{ij}=0\, , \ \  K=-\frac{d}{d-1}T\, .
\lb{13}
\ee
$T$ can be also interpreted  as the tension of the  boundary. Equation (\ref{13})  is supposed to give us a shape of boundary ${\cal S}$ provided
its own boundary $\partial{\cal S}_d=\partial{\cal M}_d$ is fixed. Inspection of this equation, however, in various situations shows that it is too restrictive
and in the absence of additional symmetries this equation is impossible to satisfy. A less restrictive condition is to impose constraint (\ref{13})
on the trace $K$ only,
\be
 K=-\frac{d}{d-1}T\, .
 \lb{14}
 \ee
 Notice that this condition alone does not follow from a variational principle.
 This is the condition which we will further  analyze.  A remark, however, should be made concerning the predictability of this prescription.
 The latter is restricted by the unknown value of parameter $T$. The other, although related,  question is what is the interpretation of $T$ from the point of view
 of the boundary CFT? Later in the paper we will  discuss  a possible answer to this question and we will relate $T$ to the certain freedom in choosing the 
 different boundary conditions in the BCFT.

Taking that the cosmological constant $\Lambda=-\frac{d(d-1)}{2}$ (we use units in which the AdS radius $l=1$) the on-shell gravitational action (\ref{12}) 
\be
W_{gr}^T=\frac{d}{8\pi G}V_{AdS}+\frac{\tanh(m)}{8\pi G}A({\cal S}_d)\, 
\lb{15}
\ee
reduces to a sum of the AdS volume and the area of the boundary ${\cal S}_d$.
Note that  we skip  the term on the boundary ${\cal M}_d$ which is not relevant to our discussion. We defined $T=(d-1)\tanh(m)$ as suggested in  \cite{Takayanagi:2011zk}
when derived 
(\ref{15}).

\subsection{Minimal surface  prescription}
We here propose an alternative prescription, motivated by the recent work on the holographic complexity \cite{Carmi:2016wjl}.
In this proposal the boundary ${\cal S}_d$  is described by embedding functions $X^\mu=X^\mu(\sigma_i)$, $\mu=1,..,d+1$ and $i=1,..,d$ so that the metric
on ${\cal S}_d$ can be written as $\gamma_{ij}(\sigma)=g_{\mu\nu}(X)\partial_i X^\mu \partial_j X^\nu$. Then we modify the gravitational action by adding a boundary volume term
\be
W^{\rm min}_{gr}=-\frac{1}{16\pi G}\int_{{AdS}_{d+1}}(R-2\Lambda)-\frac{1}{8\pi G}[\int_{{\cal M}_d}K+\int_{{\cal S}_d}\lambda]\, .
\lb{15}
\ee
The embedding functions $X^\mu(\sigma)$ are considered to be new dynamical degrees of freedom. Their values are subject to the condition 
that they describe $\partial{\cal M}_d$ when restricted to the conformal infinity of Anti-de Sitter. Variation of gravitational action with respect to
$X^\mu(\sigma)$ then gives us a condition that boundary ${\cal S}_d$ to be minimal,
\be
K=0\, .
\lb{16}
\ee
Formally, this condition corresponds to the case $T=0$ in Takayanagi's prescription. However, in gravitational action (\ref{14}) the boundary term on $\cal S$
completely vanishes when $T=0$ while in our prescription (\ref{15}) the boundary term is non-trivial even if the boundary is  minimal,
\be
W_{gr}^{\rm min}=\frac{d}{8\pi G}V_{AdS}-\frac{\lambda}{8\pi G}A({\cal S}_d)\, 
\lb{15}
\ee
It is expected, due to work of 
Graham and Witten \cite{Graham:1999pm},
that  one reproduces a conformal invariant result for the volume of a minimal surface
which bounds a subspace in the boundary of AdS.

Both the AdS volume and the area of boundary $\cal S$ are divergent. In order to regularize them we introduce a cut-off $\rho\geq \epsilon^2$.
 The holographic integral anomaly is defined  via the on-shell AdS gravitational action as follows
\be
W_{gr}=-\int_{{\cal M}_d} \la T\ra \ln\epsilon\, .
\lb{A}
\ee

\medskip

Below we will consider both prescriptions.

\section{Holographic calculation}

\subsection{Anti de Sitter metric with flat boundary}
 We consider a simple case when the boundary ${\cal M}_d$ is flat, i.e. the intrinsic Riemann tensor
vanishes identically on ${\cal M}_d$. A general form of AdS metric with such a conformal boundary takes the following form
\be
&&ds^2_{d+1}=\frac{d\rho}{4\rho^2}+\frac{1}{\rho}ds^2_d\, ,\nonumber \\
&&ds^2_d=dr^2+\gamma_{ij}(x,r)dx^idx^j\, , \nonumber \\
&&\gamma_{ij}(x,r)=\gamma^{(0)}_{ij}(x)-2k_{ij}(x)r+(k^2)_{ij} r^2\, ,
\lb{17}
\ee
where $r=0$ defines boundary $\partial{\cal M}_d$ with coordinates $x^i$, $i=1, .., d-1$. $k_{ij}$ is extrinsic curvature of the boundary defined of outward normal vector
$n^r=-1$. The form (\ref{17}) for a flat $3$-dimensional metric was earlier found in \cite{deBoer:2003vf}.
It is easy to see that $\gamma_{ij}=[\gamma^{(0)}(1-\gamma^{-1}_{(0)}k)^2]_{ij}$ is complete square. In what follows we will need its determinant,
\be
{\det}^{1/2}\gamma={\det}^{1/2}\gamma_{(0)} \det(1-\gamma^{-1}_{(0)}rk)\, .
\lb{18}
\ee
It can be easily computed as polynomial in extrinsic curvature $k$,
\be
&&\det(1-\gamma^{-1}_{(0)}rk)\equiv \alpha(r)=1+\sum_{n=1}\alpha_n r^n\, ,\nonumber \\
&&\alpha_1=-k\, , \ \ \alpha_2=\frac{1}{2}(k^2-\tr k^2)\, , \ \  \alpha_3=-\frac{1}{6}k^3+\frac{1}{2}k\tr k^2-\frac{1}{3}\tr k^3\, ,
\lb{19}
\ee
where $k=\Tr k$.
Clearly, in dimension $d$ the sum in (\ref{19}) terminates on $n=d-1$ term so that $\alpha(r)$ is polynomial in $r$ of degree $d-1$.

\subsection{ Equation for boundary $\cal S$}
 Boundary $\cal S$ is defined by the embedding function $r=r(\rho)$ such that $r=0$ if $\rho=0$.
The latter condition guarantees that $\cal S$ and ${\cal M}$ have common boundary located at $r=0$ and $\rho=0$.
Normal vector to $\cal S$ is defined as
\be
n^r=-\frac{\rho^{1/2}}{\sqrt{1+4\rho r'^2}}\, , \ \ \ n^\rho =\frac{4\rho^{3/2}r'}{\sqrt{1+4\rho r'^2}}\, , \ \ r'\equiv \partial_\rho r(\rho)\, .
\lb{20}
\ee
Equation (\ref{14}) then becomes a differential equation on $r(\rho)$,
\be
-\frac{\partial_r\alpha(r)\rho^{1/2}}{\alpha(r)\sqrt{1+4\rho r'^2}}+\rho^{(d+2)/2}\partial_\rho\left(\frac{4\rho^{-(d-1)/2}r'}{\sqrt{1+4\rho r'^2}}\right)=-\frac{d}{d-1}T\, ,
\lb{21}
\ee
solution of which determines the shape of $\cal S$. This solution can be represented as a Taylor series in $\rho^{1/2}$,
\be
r(\rho)=r_0\rho^{1/2}+r_1\rho+r_2\rho^{3/2}+r_3\rho^{2}+\dots \, .
\lb{22}
\ee
Below we will give the analysis in dimensions $d=3$ and $d=4$.

\subsection{ Dimension $d=3$}
 In three dimensions we define $T=2\tanh(m)$ in terms of another parameter, $m$. Then solving \eqref{24} order by order in $\rho$ we arrive at
\be
&&r_0=\sinh(m)\, ,  \  \  \ r_1=-\frac{\alpha_1}{4}\cosh^2(m)\, , \ \ \  r_2=\frac{1}{24}\sinh(m)\cosh^2(m)(7\alpha_1^2-16\alpha^2)\, .
\lb{d3-1}
\ee
For the found asymptotic solution  we compute the gravitational action and find that the AdS volume 
produces a logarithmic term
\be
&&V^{\rm AdS}=\int_{{\cal M}_3} d^2x\sqrt{\gamma_{(0)}}\int_{\epsilon^2}\frac{d\rho}{2\rho^{5/2}}\int_{r(\rho)}^{r_B}dr \, \alpha(r) \ , \ V^{\rm AdS}_{\rm log}=-\int d^2 x \sqrt{\gamma_{(0)}} \, {\cal V}_3\ln\epsilon^2\, , \nonumber \\
&&{\cal V}_3=\frac{\sinh(m)}{48}(-\alpha_1^2\cosh^2(m)+8\cosh^2(m)\alpha_2+8\alpha_2)\,  .
\lb{V3}
\ee
On the other hand,  the area  of boundary $\mathcal{S}$
\be
A({\cal S})=\int d^2 x \sqrt{\gamma_{(0)}} \int_{\epsilon^2}\frac{d\rho}{2\rho^2}  \sqrt{1+4\rho r'^2}  \alpha(r(\rho))\, 
\lb{d3-1-1}
\ee
produces a logarithmic term,
\be
&&A({\cal S})_{\rm log}=-\int d^2 x \sqrt{\gamma_{(0)}} {\cal A}_3 \ln\epsilon^2\, .\nonumber \\
&&{\cal A}_3=\frac{\cosh(m)}{16}\left(-2\alpha_1^2+8\alpha_2+(\alpha_1^2-8\alpha_2)\cosh^2(m)\right)\, .
\lb{d3-1-0}
\ee

\bigskip

\noindent{\it Takayanagi's prescription.} In the Takayanagi prescription this leads to a logarithm in the gravitational action (\ref{15}),
\be
W^{T}_{\rm gr, log}=-\frac{\sinh(m)}{32\pi G_N}\int_{\partial{\cal M}_3} (\hat{R}-\tr\hat{k}^2)\sqrt{\gamma_{(0)}}\, \ln\epsilon^2\, ,  \lb{d3-2}
\ee
where we used that the intrinsic curvature on $\cal S$ is $\hat{R}=k^2-\tr k^2$ and $\tr \hat{k}^2=\tr k^2-\frac{1}{2}k^2$.
So that one finds for the holographic integral conformal anomaly
\be
\int_{{\cal M}_3}\la T\ra_{\rm hol}=\frac{\sinh(m)}{16\pi G}\int_{\partial{\cal M}_3}(\hat{R}-\tr\hat{k}^2)\, .
\lb{d3-3-1}
\ee
This result is in agreement\footnote{In the first version of the paper we did not include the contribution of the AdS volume in $d=3$ and the boundary area in $d=4$ to the anomaly.
This resulted in certain discrepancies with   \cite{Miao:2017gyt} that is now fixed. We thank  Rongxin Miao for communication on this issue.}  with \cite{Miao:2017gyt}.
Comparison with (\ref{1}), (\ref{3}) shows that the boundary central charges $c_1$ and $c_2$, can be expressed in terms of parameter $m$ as follows
\be
c_1=24\sinh(m)/G\, , \ \ c_2=-16\sinh(m)/G\, ,
\lb{d3-4}
\ee
that is the ratio $c_1/c_2=-3/2$.

\bigskip

\noindent{\it Minimal surface prescription.} In the minimal surface prescription, one imposes the minimality condition, $K=0$ ($m=0$) on boundary $\mathcal{S}$.  One has in this case 
that the AdS volume (\ref{V3}) does not produce any logarithmic term while the area of boundary $\cal S$ does,
\be
{\cal A}_3=-\frac{1}{8}(\hat{R}+\tr\hat{k^2})=-\frac{1}{16}k^2\,  .
\lb{d3-6}
\ee
So that in the prescription (\ref{15}) one finds for the logarithmic term in the gravitational action,
\be
W^{\rm ms}_{gr, log}=-\frac{\lambda}{128\pi G_N}\int_{\partial{\cal M}_3} \, k^2\, \ln\epsilon^2\, .
\lb{d3-7}
\ee
Defining the trace anomaly as (\ref{A}),
the holographic integral anomaly in this prescription is
\be
\int_{{\cal M}_3}\la T\ra_{\rm hol}=\frac{\lambda}{64\pi G_N}\int_{\partial{\cal M}_3} \, k^2\, .
\lb{d3-8}
\ee
Respectively, in this case the ratio  of central charges  is $c_1/c_2=\frac{3}{2}$.
Notice that in this case $\tr k^2$ drops out in the integral trace anomaly, as is seen from (\ref{d3-6}).  This appears to be a general property of the minimal surface  prescription:
$\tr k^{d-1}$ drops out in the anomaly in dimension $d$, no matter $d$ is even or odd. For $d=4$ we will see this property in the next section.

Comparing (\ref{d3-6}) and  (\ref{d3-3-1}) we see that the two prescriptions produce different geometrical structures in the  holographic anomaly.
We also note, that in the limit $m\rightarrow 0$ the anomaly vanishes 
in the Takayanagi prescription, see (\ref{d3-2}),  (\ref{d3-3-1})  and is non-vanishing in the minimal surface prescription, (\ref{d3-7}).
Computing the anomaly on the CFT side  for a free field multiplet we could possibly  directly distinguish between the two prescriptions.

\subsection{Dimensions $d=4$}
 Following \cite{Takayanagi:2011zk} it is convenient to represent $T=3\tanh(m)$ in terms of a new parameter $m$. Then solving equation (\ref{21}) in powers of $\rho$
we determine the coefficients in the Taylor expansion (\ref{22}),
\be
&&r_0=\sinh(m)\, ,  \  \  \ r_1=-\frac{\alpha_1}{6}\cosh^2(m)\, , \ \ \  r_2=-\frac{1}{3}\sinh(m)\cosh^2(m)(\alpha_2-\frac{1}{2}\alpha_1^2)\, , \lb{23}\\
&&r_3=-\frac{1}{216}\cosh^2(m)\left(\cosh^2(m)(47\alpha_1^3-144\alpha_1\alpha_2+162\alpha_3)-40\alpha_1^3+126\alpha_1\alpha_2-162\alpha_3\right)\, ,\nonumber
\ee
where $\alpha_1$, $\alpha_2$ and $\alpha_3$ are defined in (\ref{19}).

Volume  of $AdS$ spacetime  is computed as
\be
V^{\rm AdS}=\int_{{\cal M}_4} d^{3}x\sqrt{\gamma_{(0)}}\int_{\epsilon^2}\frac{d\rho}{2\rho^3}\int_{r(\rho)}^{r_B}dr \, \alpha(r)\, ,
\lb{23}
\ee
where the integration in $r$-direction goes from the boundary $\cal S$ defined by equation $r=r(\rho)$  to some value $r_B>0$ exact value of which is not important.
Integration over $\rho$ then produces a set of divergence terms when $\epsilon$ is taken to zero. These divergences, according to the AdS/CFT dictionary
are interpreted as UV divergences. We concentrate on the logarithmic term which is found to read
\be
&&V^{\rm AdS}_{\rm log}=-\int_{\partial{\cal M}_4} \sqrt{\gamma_{(0)}}\, {\cal V}_4\, \ln\epsilon^2\, , \\
&&{\cal V}_4=(\cosh^4(m)-\frac{1}{2}\cosh^2(m))(\frac{1}{54}\alpha_1^3-\frac{1}{12}\alpha_1\alpha_2+\frac{1}{4}\alpha_3)-
\frac{1}{8}\alpha_3\, . \nonumber 
\lb{24}
\ee
On the other hand, the area of boundary $\cal S$ gives a logarithmic term
\be
&&A({\cal S})=-\int_{\partial{\cal M}_4}  \sqrt{\gamma_{(0)}} {\cal A}_4\, \ln\epsilon^2\, , \nonumber \\
&&{\cal A}_4=-\frac{\sinh^3(m)\cosh(m)}{27}(2\alpha_1^3-9\alpha_1\alpha_2+27\alpha_3)\, .
\lb{A4}
\ee
Now, substituting here values of $\alpha_k$ found in (\ref{19}) we arrive at the logarithmic term expressed in terms of 
the extrinsic curvature,
\be
&&{\cal V}_4=\frac{1}{16}Q-\frac{1}{12}(\cosh^4(m)-\frac{1}{2}\cosh^2(m))\tr \hat{k}^3\, , \nonumber \\
&&{\cal A}_4=\frac{\sinh^3(m)\cosh(m)}{3}\tr\hat{k}^3\, , \nonumber \\
&&Q=\frac{1}{3} k^3-k\tr k^2+\frac{2}{3}\tr k^3\, .
\lb{25}
\ee 
Notice that in terms of $Q$ the topological Euler number of ${\cal M}_4$ reads $\chi[{\cal M}_4]=\frac{1}{4\pi^2}\int_{\partial{\cal M}_4}Q$.
If boundary ${\cal S}$ is minimal,  $K=0$ or $m=0$, then one has
\be
{\cal V}_4=-\frac{1}{48}(k\tr k^2-\frac{5}{9}k^3)\, .
\lb{26}
\ee

\bigskip

\noindent{\it Takayanagi's prescription.}  
We find for the gravitational action (\ref{15}),
\be
W_{\rm log, gr}=-\frac{N^2}{\pi^2}\int_{\partial {\cal M}_4}\left({\cal V}_4 +\frac{\tanh(m)}{4}{\cal A}_4\right)\ln\epsilon^2\, ,
\lb{27}
\ee
where we define $N^2=\frac{\pi}{2G_N}$ according to the AdS/CFT dictionary. 
Then we find that $\cosh^4(m)$ terms are cancelled between the volume and area parts in the anomaly  and we have for the anomaly in Takayanagi's prescription,
\be
\int_{{\cal M}_4}\la T\ra_{\rm hol, T}=-\frac{N^2}{2}\chi[{\cal M}_4]+\frac{N^2}{8\pi^2}(\cosh(2m)-\frac{1}{3})\int_{\partial{\cal M}_4}\tr\hat{k}^3\, .
\lb{28}
\ee
Comparison with (\ref{4}) shows that  it correctly reproduces (for large $N$)  the $a$-anomaly in ${\cal N}=4$ super conformal gauge theory with $a=90N^2$.
On the other hand, for the boundary charge $c$ one finds agreement with (\ref{9}) provided
\be
\Delta n=70(\cosh(2m)-1)\, .
\lb{29}
\ee
Taking that in the free field approximation $\Delta n$ is an integer between $0$ and $6$, parameter $m$ has likely to take certain discret values.

\bigskip

\noindent{\it Minimal surface prescription}. Since the logarithmic divergent term in $d=4$ originates from the bulk AdS action  the holographic anomaly 
in the minimal surface prescription corresponds to (\ref{28}) for $m=0$,
\be
\int_{{\cal M}_4}\la T\ra_{\rm hol, ms}=\frac{N^2}{24\pi^2}\int_{\partial{\cal M}_4}(k\tr k^2-\frac{5}{9}k^3)\, .
\lb{30}
\ee
For large $N$ this anomaly exactly reproduces the integral anomaly (\ref{14})  in the free field approximation for the boundary conditions preserving  $1/2$ of supersymmetry, i.e. $\Delta n=0$  and $a=90 N^2$ and $c=70/3 N^2$.
Thus, the minimal surface prescription appears to be  suitable for the holographic description of ${\cal N}=4$ super-Yang-Mills theory with boundary conditions preserving $1/2$ of supersymmetry. 
Note that the boundary coupling $\lambda$ in (\ref{15})  does not appear in the anomaly in dimension $d=4$.

\subsection{Dimension $d=4$: curved ${\cal M}_4$}

In the case when space ${\cal M}_4$ is curved the calculation is more technically involved. One has to use a combination of the
Fefferman-Graham expansion in powers of $\rho$ and the expansion in powers of $\sqrt{r}$ near $\partial{\cal M}_4$. All steps of the holographic calculation
are presented in Appendix. Here we give the final result for the integral anomaly.

\bigskip

In {\it Takayanagi's prescription} the anomaly is computed to be
\be
&&\int_{{\cal M}_4}\la T\ra_{\rm hol, T}=-\frac{N^2}{2}\chi[{\cal M}_4]+\frac{N^2}{64\pi^2}\int_{{\cal M}_4} W^2_{\mu\nu\alpha\beta}+\frac{N^2}{8\pi^2}\cosh(2m)\int_{\partial{\cal M}_4}W_{n\mu n\nu}k^{\mu\nu}\nonumber \\
&&+\frac{N^2}{8\pi^2}(\cosh(2m)-\frac{1}{3})\int_{\partial{\cal M}_4}\tr\hat{k}^3\, .
\lb{31}
\ee
This is in agreement with calculation in \cite{Miao:2017gyt}\footnote{See previous footnote.}.

\bigskip

On the other hand, in the {\it minimal surface prescription} the anomaly is obtained by taking $m=0$ in previous expression, 
\be
\int_{{\cal M}_4}\la T\ra_{\rm hol, ms}=\frac{N^2}{24\pi^2}\int_{\partial{\cal M}_4}[\frac{3}{2}(k^{\mu\nu}+k n^\mu n^\nu-\frac{2}{3}kg^{\mu\nu})R_{\mu\nu}
+(k\tr k^2-\frac{5}{9}k^3)]\, ,
\lb{32}
\ee
where we dropped the bulk contributions to the anomaly and focus only on the boundary terms. We see that (\ref{32}) precisely matches  (for $N\gg 1$) the 
anomaly (\ref{10}) computed for the free super-multiplet.

\subsection{Remarks}

Let us discuss the obtained results.

\bigskip

{\bf 1.}  The main problem in using Takayanagi's prescription is how to determine parameter $T$ or, equivalently, $m$ on the CFT side of the holographic duality?
One possibility which appears to be quite natural if we look at (\ref{28}) is to associate parameter $m$ with different choices to impose the boundary conditions
in the boundary CFT. Indeed, for free fields  the boundary charge $c$ depends on the type of boundary conditions imposed on the scalars.
So that the fact that the holographic $c$ charge in (\ref{28}) depends on $m$ seems to suggest that $m$ encodes this information on the choice of the boundary
conditions as is given in (\ref{29}). However, the further inspection of other boundary terms in the anomaly (\ref{31}) indicates a problem with this
interpretation. Indeed, the boundary $b_1$ charge does not depend, for free fields, on the choice of the boundary condition. Still, in (\ref{31}), we see that
the holographic $b_1$ charge computed in Takayanagi's prescription is a function of $m$ what would be unnatural if $m$  really encoded the information
on the boundary conditions.

\bigskip

{\bf 2.} The other  issue related to the previous remark is whether the boundary charges in the anomaly are protected by the non-normalization theorems 
in the same way as charges that appear in the local conformal anomaly ($a$ and $b$)? If Takayanagi's prescription is the right one then the fact that
$b_1$ and $c$ in (\ref{31}) are non-trivial functions of $m$ should tell us that these boundary charges are not protected and may change when one switches on
the field coupling so that in the strong coupling regime they take values different from those present in the free multiplet. This, however, does not solve the problem of 
finding an intrinsic  CFT interpretation for $m$.
On the other hand, our observation  that in the minimal surface prescription $(m=0)$ the boundary charges are the same as for free fields seems to indicate that 
for those charges to be protected one needs some sufficient amount of the unbroken supersymmetry. 

\bigskip

{\bf 3.}  It is interesting to note that the effective $b_1(m)$ and $c(m)$  identified from (\ref{31}) are monotonic functions of $m$ and they take minimal values precisely 
for $m=0$. This possibly can be interpreted as some sort of monotonicity of the boundary charges with respect to the coupling strength
provided parameter $m$ somehow reflects the strength of the interaction. 
Curiously enough, $b_1(m)$ and $c(m)$ have same functionality in terms of $m$, we do not have a clear explanation for this at the moment.

\bigskip

{\bf 4.}  The observation that for $m=0$ the boundary charges do not flow and, in particular, the relation $b=b_1$ takes place in the strong coupling regime
is a strong argument in favor of the minimal surface prescription. 

\section{Conclusions}

In this paper we have presented a holographic calculation of the boundary terms in the integral conformal anomaly. We considered two prescriptions for the holographic 
description of the boundary CFT. In what we call Takayanagi's prescription the anomaly depends on an extra parameter $m$
that does not appear to have a clear physical meaning from the point of view of the boundary CFT. On the other hand, in the minimal surface prescription suggested in this 
paper this problem is absent and the holographic calculation does not contain in $d=4$ any unidentified parameter. Additional  advantage of the minimal surface
prescription is that it predicts the boundary charges to be exactly the same as in the  free field multiplet in the same way as it happens for  the bulk conformal charges
as was found in \cite{Henningson:1998gx}. We, however, are not prepared to make here a definite choice in favor of one of the prescriptions. Each prescription should pass more
tests. In particular, it would be interesting to compute the boundary entanglement entropy  holographically and reproduce  the field theory results obtained in \cite{Fursaev:2016inw}.
This work is currently in progress.

\section*{Acknowledgements} 
We thank  C. Bachas, R. Myers, K. Skenderis, M. Taylor  and M. Smolkin  for discussions and comments that influenced our work. A.F.A would like to thank M. Alishahiha, A.E. Mosaffa and A. Naseh for very useful discussions. The work of A.F.A is supported by Iran National Science Foundation (INSF).

\newpage
\appendix
\section{Details of calculation}
\setcounter{equation}0
\numberwithin{equation}{section}
We start with the following metric for AdS$_5$
\be
ds^2=\frac{1}{4\rho^2}d\rho^2+\frac{1}{\rho}g_{AB}(\rho,X)dX^AdX^B\ , \  X^A=\{r,x^i\},i=1,2,3\, .
\ee
$g_{A,B}(\rho,X)$ takes an expansion both in $r$ and  $\rho$
\be
&&g_{AB}=(1+\rho g_{rr}^{(1,0)}+r \rho g_{rr}^{(1,1)})dr^2\nn\\
&&+(g^{(0,0)}_{ij}+r g^{(0,1)}_{ij}+r^2 g^{(0,2)}_{ij}+r^3g^{(0,3)}_{ij}+\rho g^{(1,0)}_{ij}+r\rho g^{(1,1)}_{ij})dx^i dx^j\, .
\ee
where
\be
&& g_{ij}^{(0,0)}=\gamma^{(0)}_{ij}\,  ,\nn\\
& &g_{ij}^{(0,1)}=-2 k_{ij}\, ,\nn\\
&&g_{ij}^{(0,2)}=k^2_{ij}-R_{rirj}\, ,\nn \\
&&g_{ij}^{(0,3)}=-\frac{1}{3}\p_r R_{rirj}+\frac{1}{3}(k_i^\ell R_{rjr\ell}+k_j^\ell R_{rir\ell})\, ,\nn\\
&&g_{AB}^{(1,0)}=-\frac{1}{2}(R_{AB}^{(0)}-\frac{1}{6}R^{(0)}g^{(0)}_{AB})\, , \nn\\
&& g_{AB}^{(1,1)}=-\frac{1}{2}\p_r (R_{AB}^{(0)}-\frac{1}{6}R^{(0)}g^{(0)}_{AB})\, ,
\ee
where $R^{(0)}_{AB}$ and $R^{(0)}$ are constructed from the metric on the boundary of AdS, $g^{(0)}_{AB}$.\footnote{Henceforth, we drop $\cdots^{(0)}$ for simplicity.}

The components of the unit normal on the hypersurface ${\cal S}$, defined by $r=r(\rho)$, read
\be
n^r=-\frac{\rho^{1/2}(1-\rho g^{(1,0)}_{rr}-\rho r g^{(1,1)}_{rr})}{\sqrt{1-\rho g^{(1,0)}_{rr}-\rho r g^{(1,1)}_{rr}+4\rho r'^2(\rho)}}\ \ , \ \ n^\rho=\frac{4\rho^{3/2}r'(\rho)}{\sqrt{1-\rho g^{(1,0)}_{rr}-\rho r g^{(1,1)}_{rr}+4\rho r'^2(\rho)}}
\ee
Then one can compute the trace of the extrinsic curvature on ${\cal S}$ as follows
\be
K=\frac{1}{\sqrt{G}}\p_r(\sqrt{G}n^r)+\frac{1}{\sqrt{G}}\p_\rho(\sqrt{G}n^\rho)\, ,
\ee
where
\be
\sqrt{G}=\frac{1}{2\rho^3}\sqrt{g}\ \ , \ \ \sqrt{g}=\left(1+\alpha_1 r(\rho)+\alpha_2 r^2(\rho)+\alpha_3 r^3(\rho)+\beta_1\rho+\beta_2 r(\rho)\rho\right)\, .
\ee
It can be explicitly shown that
\be
\begin{split}
&\alpha_1=\frac{1}{2}\Tr g^{(0,1)} \\
&\alpha_2=\frac{1}{8}\left((\Tr g^{0,1})^2-2{\Tr g^{(0,1)}}^2+4\Tr g^{(0,2)}\right) \\
& \alpha_3=\frac{1}{48}\Big((\Tr g^{(0,1)})^3-6\Tr g^{(0,1)}{\Tr g^{(0,1)}}^2+8{\Tr g^{(0,1)}}^3\\
&-24\Tr (g^{(0,1)}g^{(0,2)})+12\Tr g^{(0,1)}\Tr g^{(0,2)}+24\Tr g^{(0,3)}\Big)\\
& \beta_1=\frac{1}{2}\Tr g^{(1,0)}\\
&\beta_2=\frac{1}{4}(\Tr g^{(0,1)}\Tr g^{(1,0)}-2\Tr(g^{(0,1)}g^{(1,0)})+2\Tr g^{(1,1)})\, .
\end{split}
\lb{beta}
\ee
Now expanding $r(\rho)$ as
\be
r(\rho)=r_0\rho^{1/2}+r_1\rho+r_2\rho^{3/2}+r_3\rho^2\, ,
\ee
and solving the equation $K=-4\tanh(m)$, one arrives at
\be
\begin{split}
&r_0=\sinh(m)\, ,\\
&r_1=-\frac{1}{6}\alpha_1\cosh^2(m)\, ,\\
&r_2=\frac{1}{6}\sinh(m)\left(\cosh^2(m)(\alpha_1^2-2\alpha_2+2\beta_1)-g^{(1,0)}_{rr}\right)\, ,\\
&r_3=\frac{1}{216}\Big(\cosh^4(m)\left(-47\alpha_1^3+144\alpha_1\alpha_2-108\alpha_1\beta_1-162\alpha_3+108\beta_2\right)\\
&+2\cosh^2(m)\left(20\alpha_1^3-63\alpha_1\alpha_2+63\alpha_1\beta_1+81\alpha_3-81\beta_2+18\alpha_1 g^{(1,0)}_{rr}\right)+27 g^{(1,1)}_{rr}\Big)\, .
\end{split}
\ee
For the minimal case where $m=0$ theses expressions reduce to
\be
\begin{split}
&r_0=r_2=0\, ,\\
&r_1=-\frac{1}{6}\alpha_1\, ,\\
&r_3=\frac{1}{216}(-7\alpha_1^3+18\alpha_1\alpha_2+18\alpha_1\beta_1-54\beta_2+36\alpha_1 g^{(1,0)}_{rr}+27 g^{(1,1)}_{rr})\, .
\end{split}
\ee
For $m=0$ the AdS volume is 
\be
\begin{split}
&V=\int_{\epsilon^2}d\rho\frac{1}{2\rho^3}\int_{r(\rho)}dr\, \alpha(r,\rho)=-{\cal V}_4\ln\epsilon^2\, ,\\
&{\cal V}_4=\frac{1}{432}(4\alpha_1^3-18\alpha_1\alpha_2+18\alpha_1\beta_1+54\beta_2-36\alpha_1 g^{(1,0)}_{rr}-27 g^{(1,1)}_{rr})\, ,
\end{split}
\ee
If $m\neq 0$ then 
\be
\begin{split}
&{\cal V}_4=-\frac{1}{8}(\alpha_3-2\beta_2)-\frac{1}{16}(\frac{4}{3}\alpha_1 g^{(1,0)}_{rr}+g^{(1,1)}_{rr})\\
&+\frac{1}{108}(\cosh^4(m)-\frac{1}{2}\cosh^2(m))(2\alpha_1^3-9\alpha_1\alpha_2+9\alpha_1\beta_1+27\alpha_3-27\beta_2)\, .
\end{split}
\ee
Substituting
\be
\begin{split}
& \Tr g^{(0,0)}=3\,  ,\\
& \Tr g^{(0,1)}=-2 k\, , \\
& \Tr {g^{(0,1)}}^2=4\Tr k^2\, , \\
& \Tr {g^{(0,1)}}^3=-8 \Tr k^3\, , \\
&\Tr g^{(0,2)}=\Tr k^2-R_{rr}\, , \\
& \Tr \left(g^{(0,1)}g^{(0,2)}\right)=-2 \Tr k^3+2k^{ij}R_{rirj}\, ,\\
&\Tr g^{(0,3)}=-\frac{1}{3}\p_r R_{rr}+\frac{4}{3}k^{ij}R_{rirj}\, , \\
&\Tr g^{(1,0)}=-\frac{1}{6}R\, ,\\
& \Tr \left(g^{(0,1)}g^{(1,0)}\right)=k^{ij}R_{ij}-\frac{1}{6}kR\, \\
& \Tr g^{(1,1)}=-\frac{1}{6}\p_r R+k^{ij}R_{ij}-\frac{1}{6}kR\, .
\end{split}
\ee
which yields
\be
\begin{split}
&\alpha_1=-k\, , \\
&\alpha_2=\frac{1}{2}(k^2-\Tr k^2-R_{rr})\, ,\\
& \alpha_3=-\frac{1}{6}\p_r R_{rr}-\frac{1}{6}(k^3-3k\Tr k^2+2\Tr k^3)-\frac{1}{3}k^{ij}R_{rirj}+\frac{1}{2}kR_{rr}\, ,\\
& \beta_1=-\frac{1}{12}R\, ,\\
&\beta_2=-\frac{1}{12}\p_r R+\frac{1}{12}kR\, .\\
& g^{(1,0)}_{rr}=-\frac{1}{2}R_{rr}+\frac{1}{12}R\, ,\\
& g^{(1,1)}_{rr}=-\frac{1}{2}\p_rR_{rr}+\frac{1}{12}\p_rR\, .
\end{split}
\ee
we can find the anomaly in terms of the curvature tensors, in particular for the minimal case we get
\be
{\cal V}_4=\frac{1}{1728}(-27\p_r R+54\p_r R_{rr}-108 k R_{rr}+36 k R+20k^3-36 k\Tr k^2)\, ,
\ee
Now using the following useful identity (see e.g. \cite{Fursaev:2015wpa})
\be
\p_r R_{rr}=\frac{1}{2}\p_r R-R_{ij}k^{ij}+kR_{rr}+T.D. \, 
\ee
we can rewrite the anomaly as
\be
{\cal V}_4=-\frac{1}{48}\left(\frac{3}{2}(k_{ij}R^{ij}+kR_{rr}-\frac{2}{3}kR)+(k\Tr k^2-\frac{5}{9}k^3)\right)\, ,
\ee
In the case of non-vanishing values of $m$ the anomaly reads in terms of the invariants
\be
{\cal V}_4&&=\frac{1}{16}Q-\frac{1}{12}(\cosh^4(m)-\frac{1}{2}\cosh^2(m)+\frac{1}{4})W_{rirj}k^{ij}\nn\\
&&-\frac{1}{12}(\cosh^4(m)-\frac{1}{2}\cosh^2(m))\Tr \hat{k}^3\, ,
\ee
where
\be
\begin{split}
&Q=R_{rirj}k^{ij}-k R_{rr}-k^{ij}R_{ij}+\frac{1}{2}kR+\frac{1}{3}k^3+\frac{2}{3}\Tr k^3-k\Tr k^2\, ,\\
&W_{rirj}k^{ij}=R_{rirj}k^{ij}-\frac{1}{2}k R_{rr}-\frac{1}{2}k^{ij}R_{ij}+\frac{1}{6}kR\, ,\\
&\Tr \hat{k}^3=\Tr k^3-k\Tr k^2+\frac{2}{9}k^3\, .
\end{split}
\ee
The area of boundary $\cal S$ is given by
\be
&&A=\int_{\epsilon^2}\frac{d\rho}{2\rho^{5/2}}\sqrt{1+4\rho g_{rr}r'^2(\rho)}\, \alpha(\rho)=-\int_{{\cal M}_4}{\cal A}_4 \ln\epsilon^2\, , \nonumber \\
&&g_{rr}=1+\rho g_{rr}^{(1,0)}+r \rho g_{rr}^{(1,1)}\, , \nonumber \\
&& \alpha(\rho)=1+\alpha_1r(\rho)+\alpha_2 r^2(\rho)+\alpha_3r^3(\rho)+\hat{\beta}_1\rho+\hat{\beta}_2\rho r(\rho)
\ee
where $\hat{\beta}_1$ and $\hat{\beta}_2$ are defined as in (\ref{beta}) but with traces defined with  respect to $3$-dimensional metric
$\gamma_{ij}$. We find
\be
{\cal A}_4=-\frac{\sinh^3(m)\cosh(m)}{27}(2\alpha_1^3-9\alpha_1\alpha_2+27\alpha_3+9\alpha_1\beta_1-27\beta_2)\, .
\ee
It can be rewritten in terms of the invariants as follows
\be
{\cal A}_4=\frac{\sinh^3(m)\cosh(m)}{3}(W_{rirj}k^{ij}+\Tr \hat{k}^3)\, .
\ee

\end{document}